\documentclass[sigconf]{acmart}

\usepackage{booktabs} 
\usepackage[english]{babel}
\usepackage[export]{adjustbox}
\usepackage{mathrsfs}
\usepackage{csquotes}
\usepackage{multirow}
\usepackage{multicol}
\usepackage{array}
\usepackage{arydshln}
\usepackage{listings}
\usepackage{booktabs}
\usepackage{subfig}
\usepackage{siunitx}
\usepackage[inline]{enumitem}
\usepackage{xspace}
\usepackage{etoolbox}
\usepackage{pifont}
\makeatletter
 
\newlist{inlinelist}{enumerate*}{1}
\setlist*[inlinelist,1]{%
  label=(\roman*),
}

\author{Luca Costa}
\affiliation{%
  \institution{Universit{\`a} della Svizzera italiana (USI)}
}
\email{luca.costa@alumni.usi.ch}

\author{Mohammad Aliannejadi}
\affiliation{%
  \institution{University of Amsterdam}
}
\authornote{Work done while Mohammad Aliannejadi was affiliated with Universit{\`a} della Svizzera italiana (USI).}
\email{m.aliannejadi@uva.nl}

\author{Fabio Crestani}
\affiliation{%
  \institution{Universit{\`a} della Svizzera italiana (USI)}
}
\email{fabio.crestani@usi.ch}

\newcommand{\app}{\texttt{Omicron}\xspace}
\newcommand{\App}{Omicron\xspace}
\newcommand{\githublink}{\url{https://github.com/aliannejadi/Omicron}}

\newcommand{\partitle}[1]{\vspace{2mm}\noindent\textbf{#1}}

\begin{document}

\title[A Tool for Conducting User Studies on Mobile Devices]{A Tool for Conducting User Studies \\ on Mobile Devices}

\begin{abstract}

With the ever-growing interest in the area of mobile information retrieval and the ongoing fast development of mobile devices and, as a consequence, mobile apps, an active research area lies in studying users' behavior and search queries users submit on mobile devices. However, many researchers require to develop an app that collects useful information from users while they search on their phones or participate in a user study.
In this paper, we aim to address this need by providing a comprehensive Android app, called \app, which can be used to collect mobile query logs and perform user studies on mobile devices. \app, at its current version, can collect users' mobile queries, relevant documents, sensor data as well as user activity and interaction data in various study settings. Furthermore, we designed \app in such a way that it is conveniently extendable to conduct more specific studies and collect other types of sensor data. Finally, we provide a tool to monitor the participants and their data both during and after the collection process.

\end{abstract}


\copyrightyear{2020} 
\acmYear{2020} 
\setcopyright{acmcopyright}
\acmConference[CHIIR '20]{2020 Conference on Human Information Interaction and Retrieval}{March 14--18, 2020}{Vancouver, BC, Canada}
\acmBooktitle{2020 Conference on Human Information Interaction and Retrieval (CHIIR '20), March 14--18, 2020, Vancouver, BC, Canada}
\acmPrice{15.00}
\acmDOI{10.1145/3343413.3377985}
\acmISBN{978-1-4503-6892-6/20/03}

\maketitle


\section{Introduction}
\label{sec:intro}

With the technological advances in developing mobile devices, we witnessed an evolution of mobile devices as well as the ways people access information through these devices. Research on mobile information retrieval (mobile IR) is aimed at ``enabling users to carry out, using a mobile device, all the classical IR operations that they were used to carry out on a desktop''~\cite{DBLP:series/sbcs/CrestaniMS17}.  The evolution of mobile devices as well as their applications (a.k.a. apps) not only enables users to carry out the ``classical IR'' tasks on their mobile devices, but also identifies new methods of interaction and information access. 

Research on mobile IR started as early as 2006 when \citet{DBLP:conf/chi/KamvarB06} studied query logs of Google mobile search. Since then, there has been growing interest in studying this area both in industry and academia. Early studies mainly focused on understanding users information needs on mobile devices~\cite{DBLP:conf/mhci/KamvarB07} and exploring conventional Web-based IR approaches on these devices~\cite{DBLP:conf/mhci/ChurchSBC08}. More recently, along with advances in technology, researchers have explored various aspects of mobile IR. For instance, \citet{DBLP:conf/www/SongMWW13} studied and found significant differences in search patterns done using iPhone, iPad, and desktop. \citet{DBLP:journals/jasis/CrestaniD06} conducted a comparative study on mobile spoken and written queries showing that spoken queries are longer and closer to natural language. \citet{DBLP:conf/chi/SohnLGH08} conducted a diary study in which they found that contextual features such as activity and time influence 72\% of mobile information needs. \citet{DBLP:conf/chi/CarrascalC15} studied user interactions concerning mobile apps and mobile search, finding that users' interactions with apps have an impact on search. \citet{DBLP:conf/sigir/HarveyP17} found that fragmented attention of users while searching on-the-go, affects their search objective and performance perception. 
Also, \citet{DBLP:conf/chiir/AliannejadiHCPC19} confirmed the findings of this study by studying people's behavior while searching in different contexts through a field study.
More recently, researchers indicated the need for a universal mobile search framework and found that commercial mobile search engines such as Google and Bing are not the preferred means of information access for the majority of users' information needs~\cite{AliannejadiSigir18}. As it is obvious, above-mentioned studies either had access to commercial search logs or developed a specific app for their study.

\begin{figure}[t]
    \centering
     \vspace{0.5cm}
        \includegraphics[width=\columnwidth]{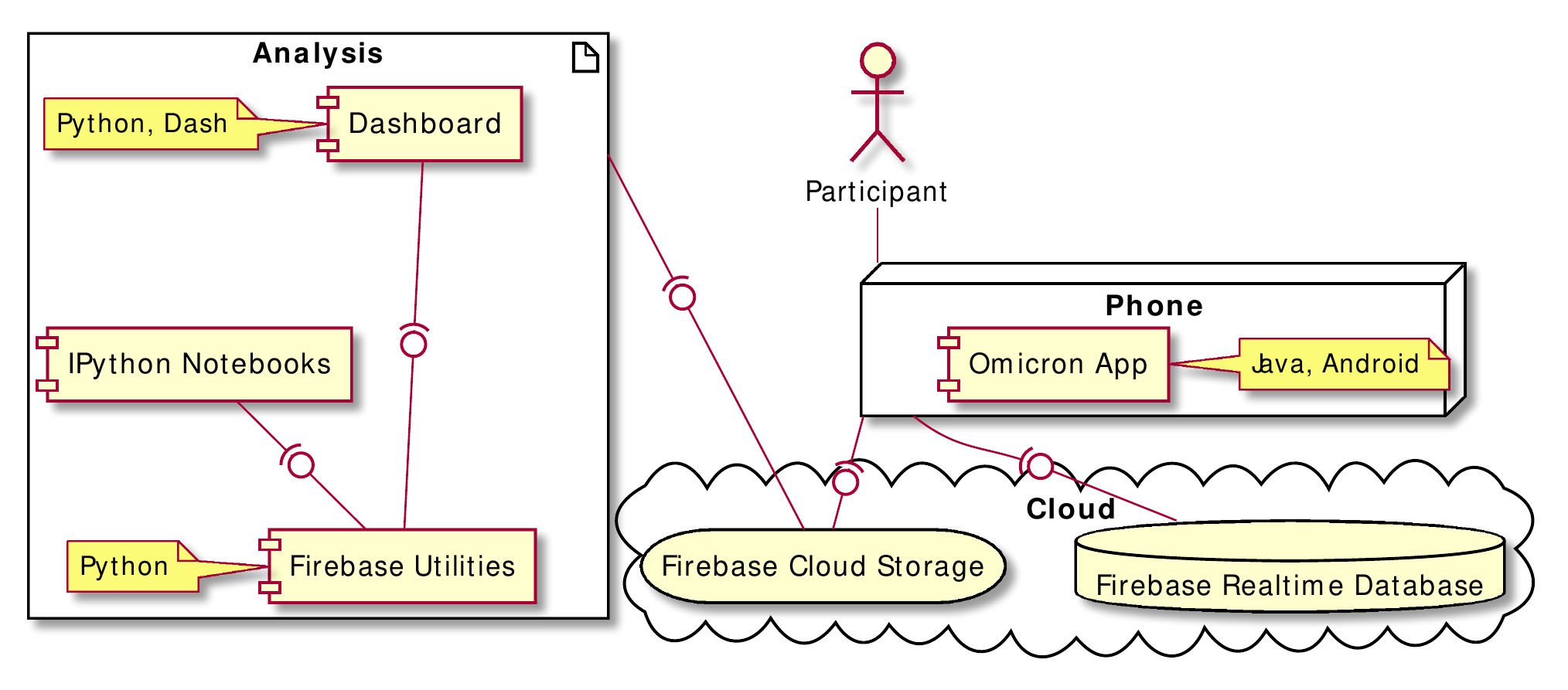}
    \caption{System Overview}
    \label{fig:overview}
\end{figure}

Due the large number of studies that have been carried out in this area, the need for a comprehensive tool that facilitates data collection in mobile IR research is clear. To this end, we have developed, tested, and open-sourced \app, to provide researchers with a tool that is comprehensive and extendable for conducting a large variety of mobile IR tasks. \app enables task-based laboratory and field studies. It can also serve as a light-weight mobile search engine allowing researchers to collect mobile query logs. Moreover, \app provides a simple way of configuring its background data collection service. The background service collects raw sensor data as well as high-level user activity information together with user interaction data while they use the app. Therefore, in this paper we aim to:
\vspace{-0.0cm}
\begin{itemize}
    \item[--] Familiarize the reader with the \app app and the monitoring dashboard.\footnote{Code available at: \githublink}
    \item[--] Give an overview of the architecture and the implementation details.
    \item[--] Elaborate on the usefulness of our system in relation to the need of an open-source customizable tool to conduct reproducible user studies.
\end{itemize}

Successful deployment and use of \app in our previous experiments indicates its usability as a tool for both conducting user studies and collecting search logs on mobile devices.

\section{\App}
Here we present the details of the user interface, collected data, and implementation of \app. 

\partitle{System overview.} An overview of the system is presented in Figure~\ref{fig:overview}. As we can see, the user interacts with \app, while the app reads and stores the data on a cloud service (i.e., Google Cloud). It is worth noting that \app stores raw sensor data, as well as user interaction data on Google Cloud Storage. On the other hand, it reads and updates the task-related data in the Firebase Realtime Database. This choice was motivated by the relative cost and effectiveness of the two services to store a large amount of data as both services does not require building and maintaining the infrastructure and offer a usage-based billing plan which make them serviceable for large-scale studies and convenient for small ones. Finally, as we see the data analysis and monitoring component retrieves the data from the cloud server and visualizes it, which could be helpful for monitoring an ongoing study.

\begin{figure*}[t]
    \centering
    \subfloat[][Tutorial]{
        \includegraphics[width=0.24\textwidth]{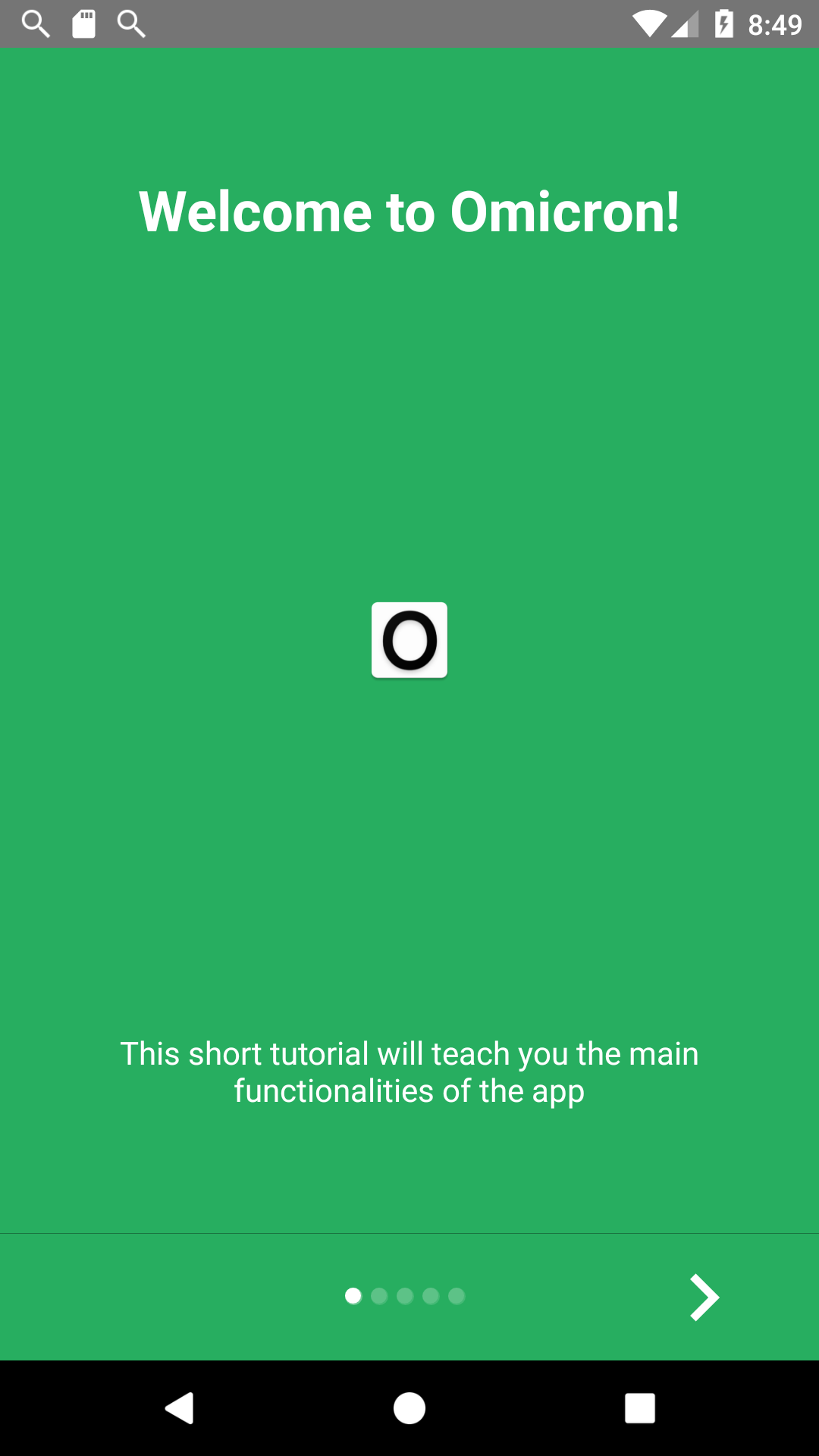}
        \label{fig:tutorial}
    }
    \subfloat[][New task]{
        \includegraphics[width=0.24\textwidth]{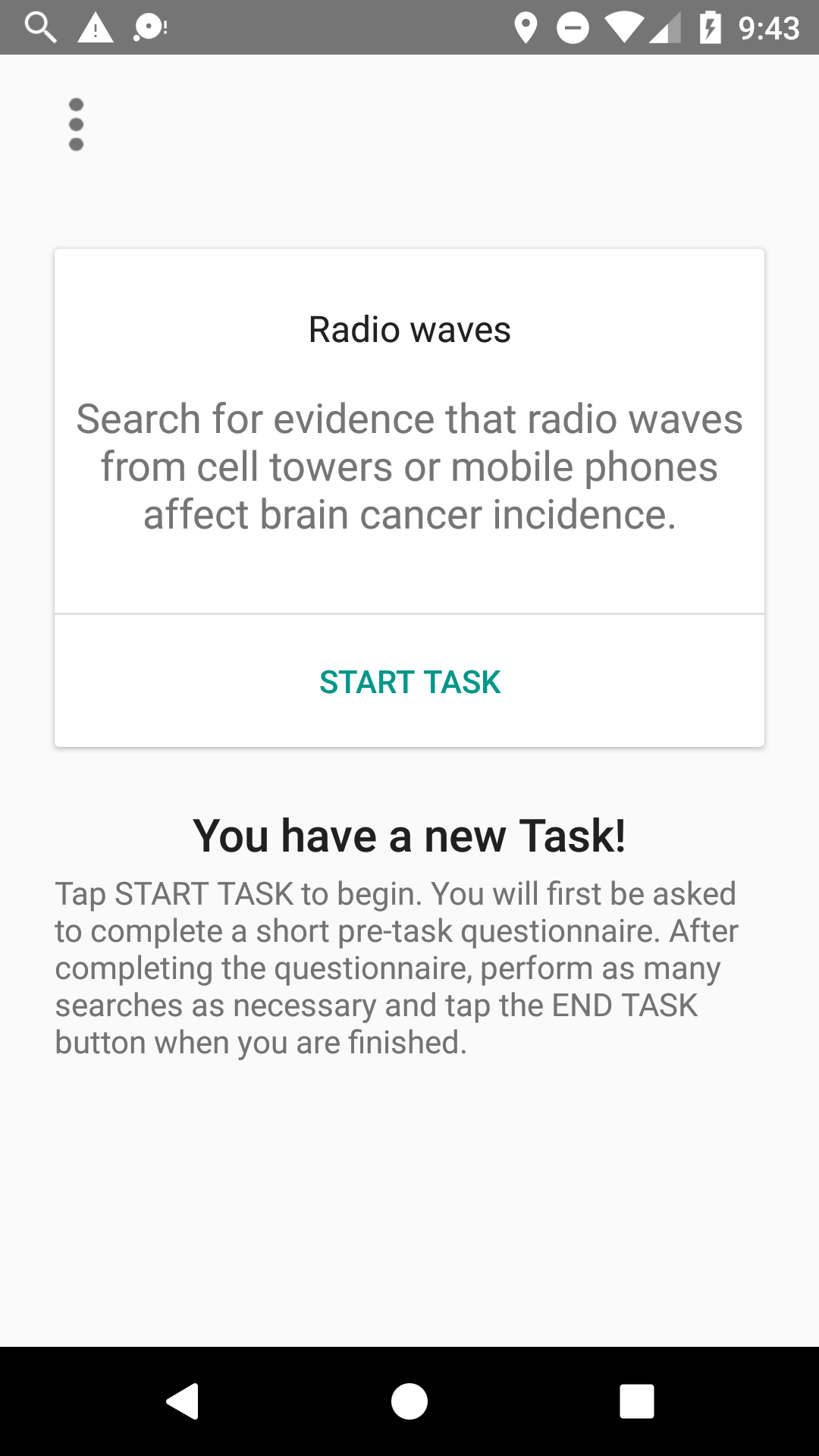}
        \label{fig:newtask}
    }
    \subfloat[][Mobile SERP]{
        \includegraphics[width=0.24\textwidth]{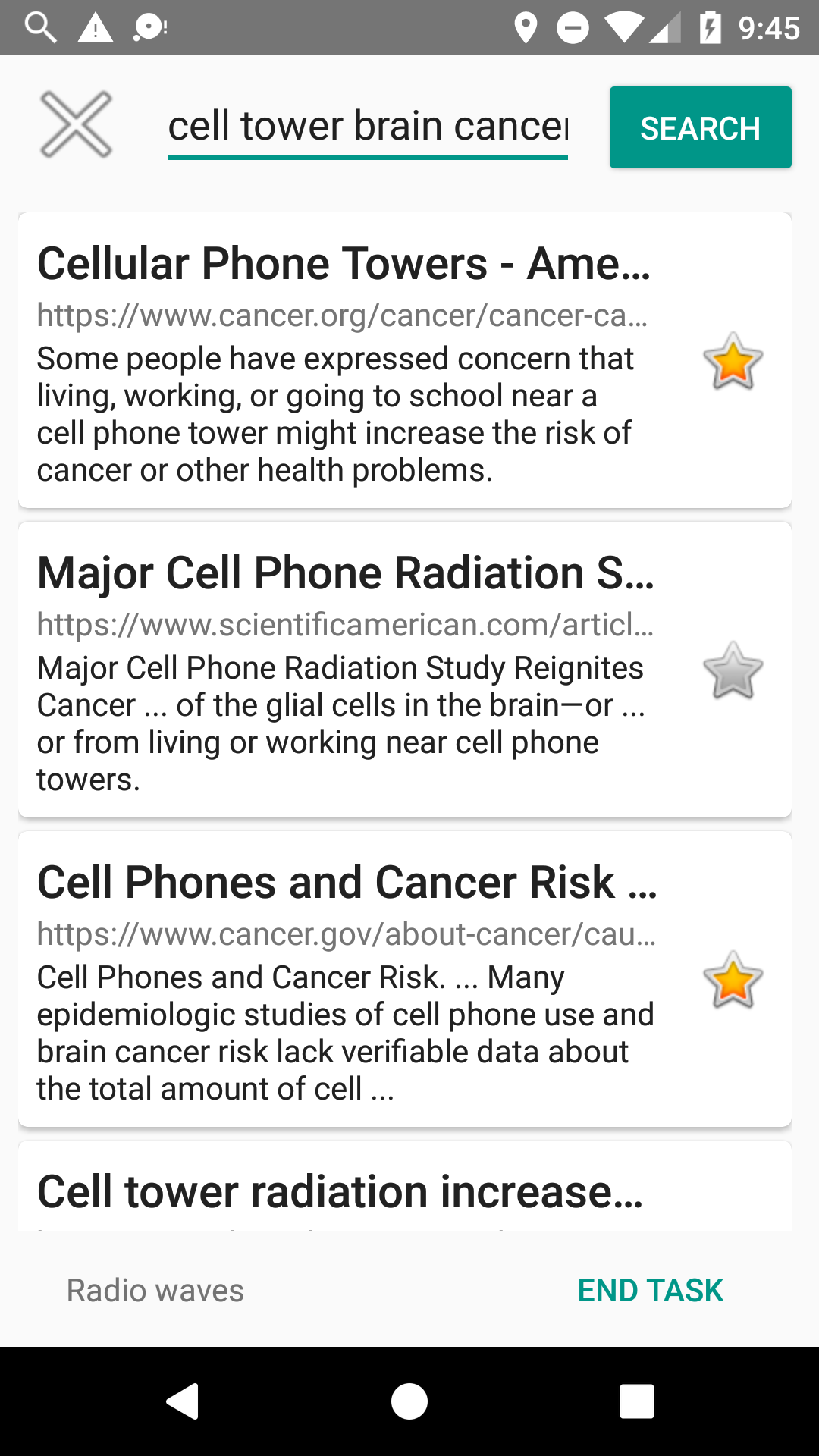}
        \label{fig:serp}
    }
    \subfloat[][Post-task Questionnaire]{
        \includegraphics[width=.24\textwidth]{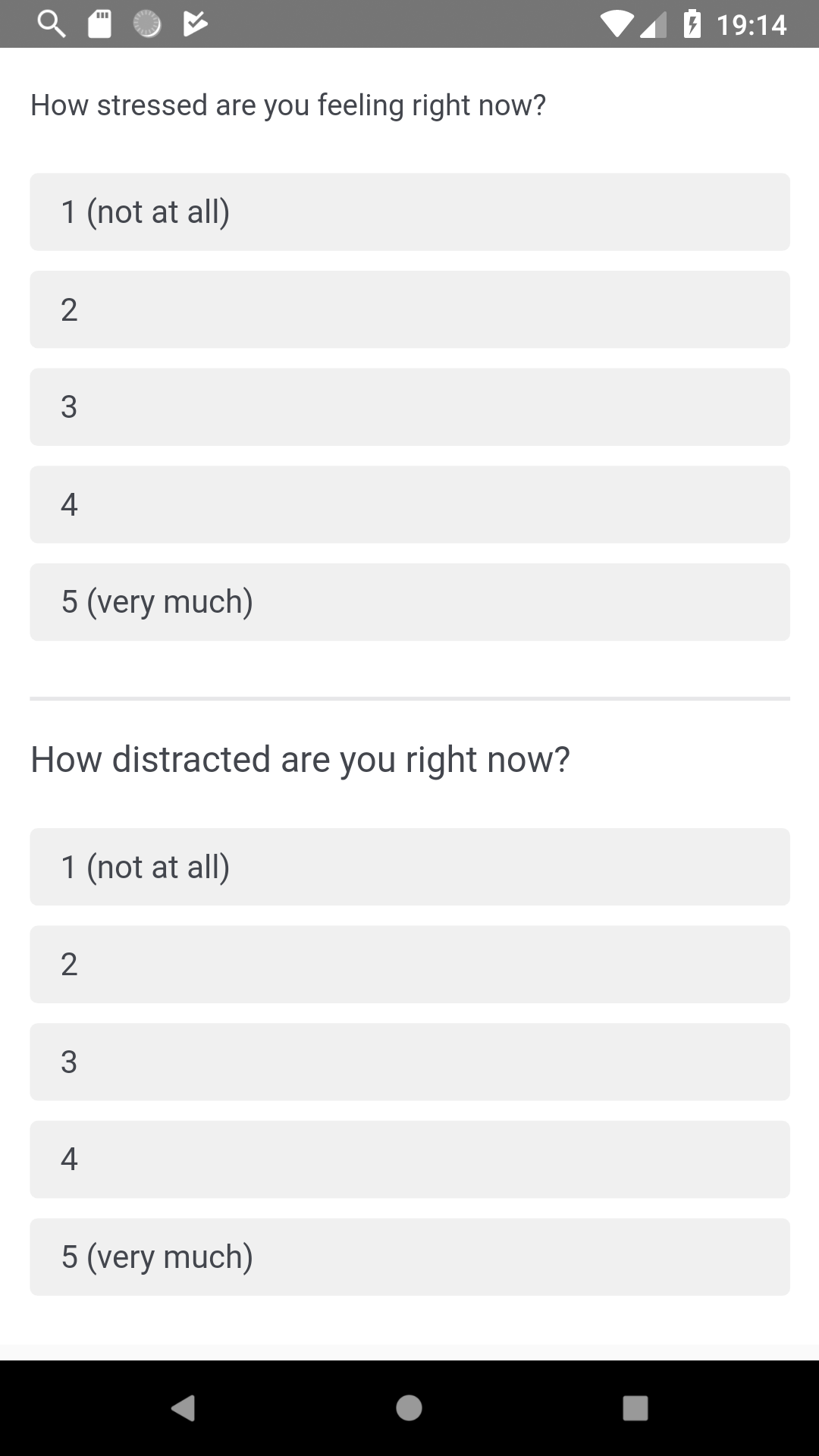}
        \label{fig:post}
    }
    \caption{Different user interfaces of \app.}
    \label{fig:task_started}
\end{figure*}

\partitle{User interface.}
\app presents itself depending on the action that the participant is required to take. After a short tutorial (Figure~\ref{fig:tutorial}), a demo task plus a demographic and background survey, participants are presented with a screen where they can freely search and test the application. The app guides the participants by showing an informative text and by hiding non-required actions at every step. When the user receives a task via a notification, the main screen changes and awaits that the participant starts the task (Figure~\ref{fig:newtask}). Figure~\ref{fig:serp} shows a sample mobile SERP where the participants consult search results and bookmark the ones that helped them complete the task. This screen is reached when the participant starts a task. Finally, as we see in Figure~\ref{fig:post}, similar to an optional pre-task questionnaire, the participants is required to complete a post-task questionnaire. 

\partitle{Collected data.} Apart from the participants' input data, \app also collects their interactions within itself~(i.e., taps and scrolling). Moreover, a background service constantly collects the phone's sensors data. In summary, \app collects the following sensor data:
\begin{itemize}
    \item GPS
    \item accelerometer
    \item gyroscope
    \item ambient light
    \item WiFi
    \item cellular
\end{itemize} 
Furthermore, it collects other available phone data that can be used to better understand a user's context. The additional collected data are as follows:
\begin{itemize}
    \item battery level
    \item the screen on/off events
    \item apps usage statistics
    \item apps usage events
\end{itemize} 
Note that apps usage statistics indicate how often each app has been used in the past 24 hours, whereas apps usage events provide more detailed app events.\footnote{\url{https://developer.android.com/reference/android/app/usage/package-summary}} 
Apps usage events record user interactions in terms of:
\begin{itemize}
    \item launching a specific app
    \item interacting with a launched app
    \item closing a launched app
    \item installing an app
    \item uninstalling an app
\end{itemize}
This service collects the data at a predefined time interval and securely transfers it to the cloud service.

\begin{table}[t]
    \begin{lstlisting}[frame=single,caption=Excerpt of the configuration file,label=configuration]
// Collectors
Bool shouldRecordAppUsage = true;
Bool shouldRecordLocation = true;
Bool shouldRecordAccelerometer = true;
Bool shouldRecordBattery = true;
        ...
// Sample rates (ms)
// app usage interval period 
long USAGE_INTERVAL = 1000*60*60*24;
// sample rate location recording
long locationRate = 1000*60*3;
// sample rate
long sampleRate = 1000 *60*2;
[...]
    \end{lstlisting}
\end{table}

\subsection{Implementation Details}
\app is created as a tool to conduct task-based user studies. It is crucial to have a way to schedule tasks, have a search interface to execute queries (in our case we used Bing) and gather sensor data at the end of the study. 

\partitle{Tasks.}
A task is defined by an ID, a start time and end time, a title, a description and a series of metadata (e.g., the time the participant start the task). For instance, we could have a task with the title "Food" and in the description specify that the participant will need to search for a recipe for dinner. We would then set a date and we would specify an interval of hours in which this task will need to be completed.

A task can be defined in various ways. For instance, each participant can have a series of tasks assigned to them on a particular day. The way tasks can be assigned to a participant or a group of them is via the application ID, an identifier generated from the instance of the installation of \app. Tasks can be changed or added during the experiment and they can be rescheduled if the participant does not manage to do them in the assigned slot.

\partitle{Configuration.}
\app can be customized by editing a single configuration file. Listing~\ref{configuration} is a small excerpt of the configuration file. It can be seen that the gathering of the various data can be conveniently switched on or off. The interval for sampling or transferring the data can also be modified easily. It is also possible to set the link for a survey after the installation and pre- and post-task questionnaires. 
Some of the sensor data can be chosen to be transmitted only over WiFi to consume less mobile data.

\begin{figure*}[t]
    \centering
        \includegraphics[width=1\textwidth]{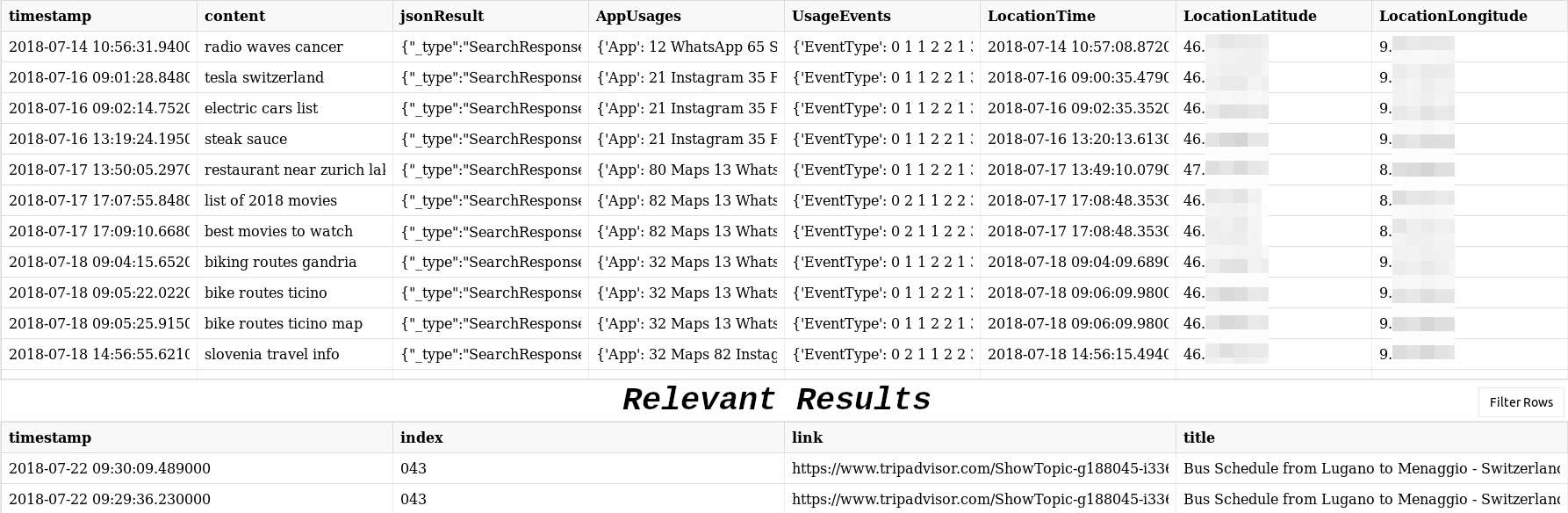}
    \caption{Part of the monitoring dashboard: The tables correlate the queries with the location, starred websites and application usage}
    \label{fig:dashboard}
\end{figure*}

\partitle{Dashboard.}
\label{sec:dashboard}
To be able to monitor the ongoing status of the study, we have built a small Web dashboard. This might be less needed for in lab studies since one can observe participants directly. In the field study setting, however, it is critical to monitor the progress of the study. We found it to be particularly useful to know when a participant was lost (the result of the queries did not answer the task) or if there was a problem in data collection. Jointly with a standard crash and error reporting dashboard, this gave us an overview of how the study was proceeding. Figure~\ref{fig:dashboard} shows one of the tables that correlated the queries with some of the other data that we collected. The dashboard is made in Python and Dash\footnote{\url{https://plot.ly/dash/}}. Although is not possible to personalize it graphically, it is relatively easy to convert Jupyter Notebooks\footnote{\url{https://jupyter.org/}}.

\section{Usability}
\label{sec:ev}
We developed \app as part of our effort to conduct an in-situ user study on smartphones. We carefully took the required measures in developing an app that runs smoothly on a large variety of Android devices. More importantly, since the participants had to install the app on their phones, we aimed to design an efficient application that does not affect the people's everyday usage experience. Therefore, we aimed to develop \app in such a way that it would not slow down the device, would not consume much energy, and would recover from possible crashes without user's involvement.

\begin{figure}
    \centering
    \includegraphics[width=\columnwidth]{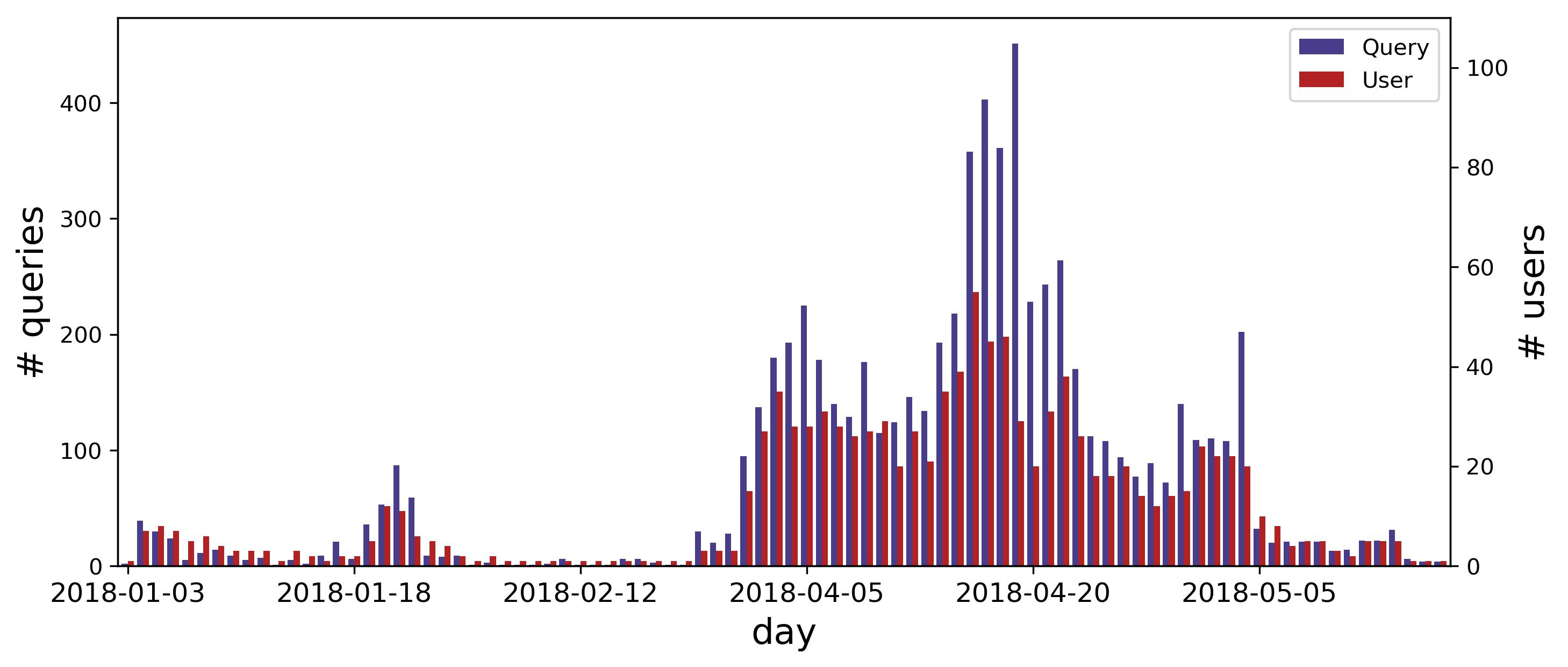}
    \caption{Number of queries and active participants per day, during the course of data collection~\cite{DBLP:conf/cikm/AliannejadiZCC18} (best viewed in color).}
    \label{fig:daily-queries}
\end{figure}

While conducting our previous studies~\cite{DBLP:conf/chiir/AliannejadiHCPC19,DBLP:conf/cikm/AliannejadiZCC18}, \app was successfully installed and used by over 350 users worldwide. Throughout the studies, we continuously monitored the collection of data, as well as the performance of the app. Several bugs were addressed throughout the study quickly, accompanied by app updates on the Play Store to ensure a positive experience of the participants. We used \app to conduct an in-situ user study where we asked the participants to perform pre-defined search tasks~\cite{DBLP:conf/chiir/AliannejadiHCPC19}. In another effort, we used the \app's background service and extended the UI to collect a mobile query search log~\cite{DBLP:conf/cikm/AliannejadiZCC18}. We see in Figure~\ref{fig:daily-queries} the number of reported queries, as well as users who installed and used the app on their phones during this study. It is worth noting that during these studies, \app collected and stored over 350 GB of data on cloud servers. In both studies, \app was installed and used on multiple smartphones from various manufacturers featuring different versions of Android OS. These successful experiences enforce the usability of \app in the community. 

Our software design allows for easy extension to collect sensor or user data that would emerge in the future. Also, the UI can be easily modified to be used for other types of studies. For instance, one could add a voice-based interface to \app for user studies based on spoken conversations. Notice that we have tested \app only for field studies and on personal phones; however, nothing limits it to be used for controlled lab studies. Therefore, one could use \app in a much more controlled experimental settings. The design of the app, as well as the database of tasks,  allows for several use cases, not to mention the possibility of extending the current version for a more specific need.

\section{Conclusions and Future Work}
\label{sec:conclusions}
In this paper we introduced \app, an app to collect mobile query logs which includes other system information about the task, and perform user studies on mobile devices. We gave a brief overview of the application and the various components that are part of the architecture. We showed how each piece fit together and how to customize \app to conduct a study. 
Writing a mobile app is expensive and time-consuming, \app offers a starting point for the researchers in the community to develop their required study apparatus on top of \app, without worrying about a bespoke application. 
Furthermore, we have designed \app to be easy to customize so that it can adapt to the needs of different studies. It can also be easily configured to perform other types of studies, or be used as a simple tool to collect mobile query log.
Finally, it is worth noting that \app provides the ground towards more reproducible studies.

In the future, we plan to extend \app to be modular and explore gamification for mobile data collection. One of the challenges of building an application that can be reused for different types of user studies is that for instance, one study might need to gather sensor data but it would not make use of the search feature. Unfortunately at the moment, it is hard to do it without adventuring in the code base and knowing Android programming. The architecture could be made more modular to allow both the reuse of low-level components and graphical components, doing this could allow easier reuse of \app even for studies that differ considerably from what it was created for.
As we know, it is difficult to gather enough participants to conduct mobile IR studies at large scale. By making the studies open via mobile app stores it is possible to sample from a vast audience as shown in studies like \cite{DBLP:conf/nordichi/HenzePB10} and \cite{mcmillan2010further}. Every study would need to adapt game elements in a way that incentivize
participants to take part in their studies. Therefore the type of study would dictate a different approach. However, there are known game elements that can be reused in many studies, for instance leader boards and points. Also, we plan to extend \app to enable studies on conversational search on mobile devices~\cite{DBLP:conf/sigir/AliannejadiZCC19}.

\begin{acks}
    This research was in part funded by the RelMobIR project of the \grantsponsor{}{Swiss National Science Foundation (SNSF)}{http://www.snf.ch/en/Pages/default.aspx}. We would like to thank Jacopo Fidacaro, who helped us in developing \app as part of his summer internship. Also, we would like to thank Morgan Harvey and Matthew Pointon for their feedback on the app's design.
\end{acks}

\Urlmuskip=0mu plus 1mu\relax
\bibliographystyle{ACM-Reference-Format}
\bibliography{sigproc} 

\end{document}